\documentclass[12pt,a4paper,british]{iopart}
\usepackage{amssymb}
\usepackage{amsgen}
\usepackage{bbold}
\usepackage{color}
\usepackage[T1]{fontenc}
\usepackage[latin9]{inputenc}
\usepackage{babel}
\usepackage{graphicx}
\usepackage{epstopdf}
\usepackage[numbers,square,sort&compress]{natbib}
\usepackage{array}
\usepackage{verbatim}
\usepackage{amstext}
\usepackage{iopams}
\expandafter\let\csname equation*\endcsname\relax
\expandafter\let\csname endequation*\endcsname\relax
\usepackage{amsmath}
\usepackage{amsfonts}
\usepackage{float}

\makeatletter

\renewcommand{\t}[1]{\textrm{#1}}
\newcommand{\ot}{\otimes}

\newcommand{\bra}[1]{\langle #1|}
\newcommand{\ket}[1]{|#1\rangle}

\newcommand{\proj}[1]{|#1\rangle \langle #1|}
\renewcommand{\t}[1]{\textrm{#1}}
\newcommand{\norm}[1]{\|#1\|}
\newcommand{\rank}{\textrm{rank}}
\newcommand{\hmin}[2]{H_{\text{min}}(#1)_{#2}}
\newcommand{\hmax}[2]{H_{\text{max}}(#1)_{#2}}
\newcommand{\hemin}[2]{H^{\epsilon}_{\text{min}}(#1)_{#2}}
\newcommand{\hemax}[2]{H^{\epsilon}_{\text{max}}(#1)_{#2}}

\begin{document}
\title{Thermodynamic work cost of quantum estimation protocols}
\author{Patryk Lipka-Bartosik$^{1,2}$, Rafa{\l} Demkowicz-Dobrza{\'n}ski$^1$}
\address{$^1$ Faculty of Physics, University of Warsaw, Pasteura 5, PL-02093 Warsaw, Poland}
\address{$^2$ H. H. Wills Physics Laboratory, University of Bristol, Tyndall Avenue, Bristol, BS8 1TL, United Kingdom}

\begin{abstract}
We discuss thermodynamic work cost of various stages of a quantum estimation protocol: probe and memory register preparation,
measurement and extraction of work from post-measurement states. We consider both (i) a multi-shot scenario, where average work is calculated in terms of the standard Shannon entropy and (ii) a single-shot scenario, where deterministic work is expressed in terms of min- and max-entropies.
We discuss an exemplary phase estimation protocol where estimation precision is optimized under a fixed work credit (multi-shot) or a total work cost (single-shot). In the multi-shot regime precision is determined using the concept of Fisher information, while in the single-shot case we advocate the use of confidence intervals as only they can provide a meaningful and reliable information in a single-shot experiment, combining naturally with the the concept of deterministic work.
\end{abstract}
\maketitle

\section{Introduction}
Since the formulation of the Landauer principle \cite{Landauer1961} and the resolution of the infamous paradox of Maxwell's Demon \cite{Maxwell1871} by Bennett \cite{Bennett1982}, Shannon entropy ceased to be a purely informational concept and found its well deserved place in the realm of physics. In his analysis, Bennett realized that although measurement itself may in principle be performed without spending work,
removal of each bit of information stored in Demon's mind requires at least $k_B T \ln 2$ work. This is also the maximal amount of work that could be drawn if the Demon used his knowledge about measurement results to extract work from the system. Hence, if the total state of the system \emph{and} the Demon is considered, then either the entropy of the Demon's mind increases and hence the work is not given for free, or the state of Demon's mind resets to its original state at the expense of work. This is the basic idea of how the apparent paradox is resolved.

In short, the above analysis deals with the  relation between information and work. In this paper, in a similar spirit, we investigate the relation between work cost of an estimation protocol and precision of measuring an unknown parameter encoded in a quantum state. Moreover, unlike in the canonical analysis of the Maxwell's Demon, we do it in a fully quantum paradigm. On one hand, the information is obtained from a result of a quantum measurement performed on a quantum system, as in the paradigm of quantum estimation theory \cite{Helstrom1976, Holevo1982, Giovannetti2006, Toth2014, Demkowicz2015, Dorner2009, Demkowicz2009, Mazzola2013, Kolodynski2010, Smirne2016, Nichols2017}. On the other hand, the thermodynamic cost of the estimation procedure is quantified formally using quantum resource theories recently developed within the field of quantum thermodynamics \cite{Horodecki2013, Goold2016, Strasberg2017, Faist2018, Skrzypczyk2014, Winter2016, Oppenheim2002, Sparaciari2017, Linden2009, Popescu2018, Faist2015gpm, Brandao2015, Guryanova2016, Huber2015, Chiribella2017, Lostaglio2015, Misra2016}.

Here we discuss all relevant stages of an estimation protocol and assign appropriate work costs for each of them. Furthermore, we clearly distinguish between single and multi-shot regimes. It is a crucial distinction, both in the estimation theory and quantum thermodynamics. In estimation theory, both classical and quantum, a scenario where multiple repetitions of an experiment are available
leads to a huge simplification of the problem. In such scenarios one can easily determine the optimal measurements and estimators which, in the limit of large number of repetitions, saturate the Cram{\'e}r-Rao bound and hence are proven to be optimal. Similarly in quantum thermodynamics, the multiple copy scenarios in the asymptotic regime reproduce results known from classical thermodynamics. It is therefore particularly interesting to focus on single-shot scenarios where, on one hand, we draw conclusions regarding the parameter of interest from a single measurement outcome and, on the other hand, we want to be sure that after fixing a desired precision, a given amount of deterministic work is sufficient to perform such a protocol. 

Some aspects of the relation between quantum estimation precision and its thermodynamics cost have been addressed by other authors using different perspectives on the problem and focusing on different goals \cite{Micadei2013, Scorpo2018, Erker2017}.
In \cite{Micadei2013} authors focused solely on the preparation stage of quantum estimation protocols in the many-copy regime. 
In \cite{Erker2017} the main focus was on studying the trade-off between the precision of a clock and its stability vs. the read-out process
in a models where the clock operates as an autonomous machine powered by thermal baths. 
Finally, in \cite{Scorpo2018}, the authors considered a more complicated metrological model (multiple-qubits, noise correlations), but they took a simpler approach to quantification of work/energy aspects of the protocol. They focused solely on the changes of mean energy and considered only unitary transformations that can be applied on the probe systems, so as a result the entropy of the state remained the same throughout the protocol.
 
Here we address a broader question of the total cost of a general quantum estimation protocol and make a clear distinction between the multi-shot and single-shot regimes, offering in each regime a clear connection between the relevant estimation and thermodynamic concepts.
It is important to point out that we will see a fundamental irreversibility of the protocol in the single-shot scenario and hence, in this regime, we will address a real work cost of the whole protocol. However, we will also realize that in the multi-shot regime all steps may be in principle performed in a reversible way and thus the net cost of the protocol vanishes. Still, we may ask what is the \emph{work-credit}, that is, the amount of work we need to invest during the protocol. We will treat this quantity as a resource which limits our ability to estimate the true value of the parameter in the multi-shot regime.

The paper is organized as follows. In Section~\ref{sec:thermo} we review recent results on the work cost of quantum operations from the perspective of quantum resource theories. For simplicity of presentation, we focus here on the degenerate Hamiltonian case, which is sufficient to capture the essence of the relations between metrological and thermodynamical aspects of the estimation protocols.  In Section~\ref{sec:protocol} we discuss a general quantum parameter estimation protocol and
distinguish three different stages relevant for determining the work cost. In Section~\ref{sec:workcostprotocol} we combine results from two previous sections to give explicit formulas for work costs of different stages of the estimation protocol, both in the single and multi-shot regimes. In Section~\ref{sec:example} we illustrate these concepts using a simple single-qubit estimation problem, which allows us to easily optimize estimation performance for a fixed work cost in both regimes. Finally in Section~\ref{sec:nondeg} we will generalize our consideration to the non-degenrate Hamiltonian case and also restudy the single-qubit protocol from Section~\ref{sec:example} within this more general framework. We conclude the paper in Section~\ref{sec:conclusions}.

\section{Work cost of quantum operations}
\label{sec:thermo}
In order to find thermodynamic work associated with the estimation protocol we adapt a well-established framework for quantum thermodynamics called \emph{resource theory of Gibbs-preserving maps} \cite{Goold2016,Korzekwa2016,Faist2015,Faist2018,Strasberg2017}. Free operations of this theory are completely positive and trace-preserving maps $\Phi_{X \rightarrow X'}$ acting between input system $X$ and output $X'$ which preserve the Gibbs state $\tau_X =  e^{-\beta \mathcal{H}_X} / Z_X$, where $\beta$ is an inverse temperature, $\mathcal{H}_X$ is the Hamiltonian of system $X$ and $Z_X = \tr e^{-\beta \mathcal{H}_X}$ is the associated partition function. In what follows, for the sake of clarity and simplification of the presentation, we will mostly focus on a simplified model of thermodynamics in which Hamiltonian of the system and memory register are fully degenerate, meaning that their respective thermal states are maximally mixed states $\tau_X = \mathbb{1}_X / |X|$, where $|X| = \dim(X)$ (once we get a proper understanding of this simplified case, we will discuss the more general non-degenerate Hamiltonian case in Section~\ref{sec:nondeg}).
This means that the condition of preserving Gibbs state simplifies to:
\begin{equation}
    \label{eq:2}
    \Phi_{X\rightarrow X'} \left[ \mathbb{1}_X\right] = \mathbb{1}_{X'}.
\end{equation}
If transformation is not of the form (\ref{eq:2}) then we say that it is a non-free operation. Operations which are non-free can still be performed, however, the agent needs to supply a certain amount of thermodynamic work to execute them. It is a long standing question of how to account for work in the quantum regime. Here we take an operational point of view and adapt a very promising idea of an \emph{information battery}. This model dates back to Bennet \cite{Bennett1982} and Feynman \cite{Feynman1998}, and very recently was successfully used by Faist to obtain bounds on the minimal work cost of implementing any quantum channel \cite{Faist2015,Faist2018}. Such a battery is a system which consists of a large number $D$ of qubits ($D \gg 1$) with a degenerate Hamiltonian and each of which can be prepared either in a pure state $\ket{0}$ or a maximally-mixed state $\mathbb{1}_2/2$. Denoting the battery system with $A$ we can write an arbitrary state of the battery with $l \leq D$ maximally-mixed qubits as:
\begin{equation}
    \label{eq:3}
    \rho_A(l) = \proj{0}^{\otimes (D-l)} \otimes \frac{1}{2^l} \mathbb{1}_l.
\end{equation}
The general idea behind this battery model is based on the fact that a pure qubit can be used to perform $k_B T \ln 2$ of deterministic work (by using a Szilard box \cite{Szilard1964}) or analogously, a maximally mixed qubit can be transformed back into a pure state deterministically at the same work cost using Landauer erasure \cite{Landauer1961}. If an agent wants to perform a non-free transformation $\mathcal{C}_{X\rightarrow X'}$ which takes $\rho_X$ to $\rho_{X'}$, that is $\mathcal{C}_{X\rightarrow X'}[\rho_X] = \rho_{X'}$, then she can start with the battery in state $\rho_A(l)$ and consider a free process $\Phi_{XA \rightarrow X'A}$. This process acts on the joint state $\rho_X \otimes \rho_A(l)$ and performs the action of $\mathcal{C}_{X\rightarrow X'}$ on system $X$, while taking the battery to a state with $l^\prime$ maximally mixed qubits. The joint transformation can be written as:
\begin{equation}
     \label{eq:4}
    \Phi_{XA \rightarrow X'A} \left[ \rho_X \otimes \rho_A(l) \right] = \mathcal{C}_{X\rightarrow X'} \left[\rho_X\right] \otimes \rho_A(l^\prime).
\end{equation}
The amount of work consumed by the process $\mathcal{C}_{X\rightarrow X'}$ is given by the difference $k_B T \ln 2 \times (l^\prime- l)$. In what follows, whenever we discuss the amount of work, we will drop  the $k_B T \ln 2$ coefficient, so in fact all work quantities are given in terms of this unit.  Sometimes the process can also exploit purity of the input state $\rho_X$ to increase the total number of pure qubits inside the battery. In this situation $(l^\prime - l)$ is negative and can be interpreted as the work extracted by $\mathcal{C}_{X\rightarrow X'}$ from the state $\rho_X$. 

A very elegant result by Faist and Renner \cite{Faist2015} states that
for any channel $\mathcal{C}_{X\rightarrow X'}$ there exist battery states with $l$ and $l^\prime$ maximally mixed qubits such that
there exist a free operation $\Phi_{XA \rightarrow X'A}$ such that (\ref{eq:4}) holds. Moreover, the authors found that the minimal number of pure qubits which must be consumed (or the maximal number which can be extracted from the input state) in order to perform channel $\mathcal{C}_{X\rightarrow X'}$ reads
\begin{equation}
    \label{eq:5}
    w(\mathcal{C})  = l^\prime - l = \log \norm{\mathcal{C}_{X\rightarrow X'} \left[ \Pi_{\rho_X} \right]}_{\infty},
\end{equation}
where $\Pi_{\rho_X}$ is a projector onto the support of the input state $\rho_X$, $\norm{\cdot}_{\infty}$ is the infinity norm and $\log$ is (and will be throughout this paper) implicitly taken as the logarithm with base $2$.

Before we proceed further, let us introduce some information-theoretic quantities which allow to express (\ref{eq:5}) in a more illustrative way. First, recall the definition of the \emph{Shannon entropy} of a probability distribution $\{p_x\}$ as well as the \emph{von Neumann} entropy of a quantum state $\rho_X$:
 \begin{equation}
 H(\{p_x\}) = -\sum_x p_x \log p_x, \quad  H(\rho_X) = H(X)_\rho =  -\tr \rho_X \log \rho_X,
 \end{equation}
which coincide if $\{p_x\}$ are eigenvalues of $\rho_X$, $\rho_X = \sum_x p_x \proj{x}$.
The conditional variant of the von Neuman entropy for a joint state $\rho_{XY}$ of two quantum systems $X$ and $Y$ reads:
  \begin{equation}
  H(X|Y)_{\rho} := H(XY)_{\rho} - H(Y)_{\rho},
   \end{equation}
and unlike its classical counterpart can also be negative \cite{Horodecki2005}.

Shannon and von Neuman entropies are operationally meaningful quantities in many information-theoretic tasks such as communication, information compression or randomness extraction. However, this is true only when the task is performed in the so-called \emph{multi-shot} regime, that is, when protocols operate on many independent entities, be it quantum states, channels or random variables.
   If, however, the task is to be carried out only once, then meaningful information is  provided by different entropy measures. This regime is known as the \emph{single-shot} regime and involves many diverse (and often very specialized) entropy measures \cite{Renyi1961,Renner2008, Renner2004}. A readable review of these measures and their properties can be found in \cite{Tomamichel2015}.

In this paper we will make extensive use of the so-called \emph{min-} and \emph{max-entropies} which are defined, respectively, as:
    \begin{equation}
    H_{\text{min}}(X)_{\rho} := - \log \norm{\rho_X}_{\infty}, \quad H_{\text{max}}(X)_{\rho} :=  \log \Tr \Pi_{\rho_X}  = \log \rank \, \rho_X.
    \end{equation}
If side information is available in another system $Y$, then single-shot tasks in general can be performed more efficiently. This is captured by \emph{min-} and \emph{max-conditional entropies} defined by:
\begin{eqnarray}
    \hmin{X|Y}{\rho} &:= - \log \min_{\sigma_Y} \{ \tr \sigma_Y \,|\, \mathbb{1}_X \ot \sigma_Y \geq \rho_{XY} \}, \\
    \hmax{X|Y}{\rho} &:= \log \norm{\tr_X \Pi_{\rho_{XY}}}_{\infty}
\end{eqnarray}
and similarly to conditional von Neuman entropy can be negative for entangled states \cite{Delrio2011,Tomamichel2015,Faist2016}. Note that
for simplicity of further calculations, we use the variant of max-conditional entropy which is often referred to as R{\'e}nyi zero conditional entropy---see \cite{Faist2015} for further discussions. Importantly, notice that conditional min- and max-entropies in general cannot be expressed as differences of appropriate unconditional quantities as is the case with von Neuman and Shannon entropies.

In order to establish a connection between singe-shot entropies and the standard notion of entropy (and to make these quantities continuous functions of the state) it is necessary to introduce the concept of smoothing. Following \cite{Renner2008}, the \emph{conditional smooth entropies} for a smoothing parameter $0 < \epsilon < 1$ are defined as
\begin{eqnarray}
\hemin{X|Y}{\rho} &:= \max \{ \hmin{X|Y}{\tilde{\rho}} \, |\, {\tilde{\rho} \approx_{\epsilon} \rho } \}, \\
\hemax{X|Y}{\rho} &:= \min \{ \hmax{X|Y}{\tilde{\rho}} \, |\, {\tilde{\rho} \approx_{\epsilon} \rho } \},
\end{eqnarray}
where  $\tilde{\rho}_X \approx_{\epsilon} \rho_X$, means that $\sqrt{1-F^2(\rho, \tilde{\rho})} \leq \epsilon$, where $F(\rho, \sigma)$ is the fidelity between $\rho$ and $\tilde{\rho}$. The above quantities earn their operational meaning when one considers single-shot tasks that are allowed to fail with some small probability $\epsilon \ll 1$.

The asymptotic equipartition property \cite{Tomamichel2015} implies that for any $0 < \epsilon < 1$ smoothed entropies calculated on a quantum state $\rho_{XY}^{\ot n}$ representing many ($n \gg 1$) independent and identically distributed copies of a quantum state $\rho_{XY}$, approach asymptotically the standard von Neuman entropy:
\begin{equation}
\lim_{n \rightarrow \infty} \frac{1}{n}\hemin{X|Y}{\rho^{\otimes n}} = \lim_{n \rightarrow \infty} \frac{1}{n}\hemax{X|Y}{\rho^{\otimes n}} = H(X|Y)_{\rho}.
\end{equation}
This establishes a connection between single-shot quantities calculated on a single representative state of a many-copy sample and the multi-shot quantities inferred from the sample.

Following \cite{Faist2016}, the single-shot work cost of a quantum channel  $\mathcal{C}$ as given by equation (\ref{eq:5}) can also
be written in terms the max-entropy:
\begin{equation}
    \label{eq:6}
    w(\mathcal{C}) =  \hmax{E|X'}{\rho_{X'RE}}.
\end{equation}
In the above formula $R$ is a reference system used to purify the input state $\rho_X$ to a pure state $\rho_{XR}$, such that $\tr_R [\rho_{XR}] = \rho_X$.
The action of $\mathcal{C}_{X\rightarrow X'}$ extended trivially to $R$ yields state $\rho_{X'R} = \left(\mathcal{C}_{X\rightarrow X'} \otimes \mathbb{1}_R \right)\left[\rho_{XR}\right]$ which contains full information about the channel $\mathcal{C}_{X\rightarrow X'}$ and  the input state $\rho_X$. Finally, $E$ is the environment system used to write a unitary dilation of $\mathcal{C}_{X \rightarrow X'}$, meaning that $\mathcal{C}_{X \rightarrow X'}[\rho_X] = \tr_{RE}[\rho_{X'RE}]$ with $\rho_{X'RE} = U \rho_{XR} \ot \ket{0}\bra{0}_E  U^\dagger$ and $U$ beeing a unitary dilation of $\mathcal{C}_{X \rightarrow X'}$ trivially extended to $R$.
We can obtain the formula for the \emph{multi-shot work cost} $\langle w(\mathcal{C}) \rangle$ as the multi-copy limit of the smoothed version of (\ref{eq:6}), that is:
\begin{eqnarray}
\label{eq:workmulti}
    \langle w(\mathcal{C}) \rangle
    &=  \lim_{n \rightarrow \infty} \, \frac{1}{n} \, \cdot \hemax{E|X'}{\rho_{X'RE}^{\ot n}} \\ \nonumber
    &= H(E|X')_{\rho_{X'RE}} \\ \nonumber
    &= H(E X')_{\rho_{X'RE}} - H(X')_{\rho_{X'RE}} \\ \nonumber
    &=  H(R)_{\rho_{X'RE}} - H(X')_{\rho_{X'}}\\  \nonumber
    & = H(R)_{\rho_{X R}} - H(X^\prime)_{\rho_{X'}} \\ \nonumber
    &= H(\rho_X) - H(\rho_{X^\prime}),
\end{eqnarray}
where we used the fact that both $\rho_{X^\prime R E}$ and $\rho_{XR}$ are pure. We see that taking the multi-copy limit recovers the familiar thermodynamic formula where work is given by the change of free energy, which in this case (degenerate Hamiltonian) coincides with the change of the von Neumann entropy.

We emphasise that the single-shot variant given in (\ref{eq:6}) generally cannot be written as a difference of entropies of the input and output states.
Still, in a situation where channel $\mathcal{C}_{X \rightarrow X'}$ produces a $\emph{fixed}$ output state $\rho_X^\prime$ irrespectively of the input $\rho_X$, formula (\ref{eq:5}) simplifies to:
\begin{eqnarray}
\label{eq:workfixedout}
\nonumber
w(\mathcal{C}) &= \log \norm{\rank{\, \rho_{X}}\, \cdot\, \mathcal{C}_{X \rightarrow X'} \left( \Pi_{\rho_X} / \rank{\, \rho_{X}} \right)}_{\infty}  \\ \nonumber
&= \log \left( \rank \, \rho_X \cdot \norm{\rho_X'}_{\infty} \right) \\ &=  \hmax{\rho_X}{} - \hmin{\rho_{X^\prime}}{},
\end{eqnarray}
a formula which will come useful when discussing the work cost of quantum estimation protocols in Section~\ref{sec:workcostprotocol}.

\section{Quantum parameter estimation protocol}
\label{sec:protocol}
\begin{figure}
    \centering
    \includegraphics[scale=0.4]{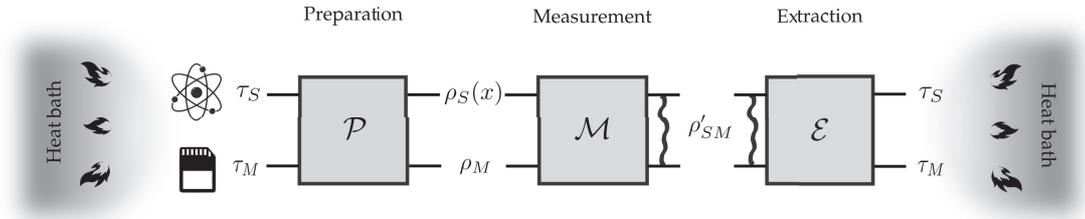}
    \caption{General scheme of a quantum parameter estimation protocol. In the first step of the protocol the agent (Alice) draws two thermal states from a heat bath (system $\tau_S$ and memory $\tau_M$). Using channel $\mathcal{P}_{SM \rightarrow SM}$ she prepares
    the probe state $S$ on which information about a parameter $x \in \mathbb{R}$ is encoded resulting in a state $\rho_S(x)$ and erases memory register to a state $\rho_M$ (possibly mixed). In the next step she performs a general quantum measurement using channel $\mathcal{M}_{SM \rightarrow SM}$ and obtains a classically correlated state of the system and memory $\rho_{SM}'$, where the information encoded in $M$ is the basis for the estimation procedure. In the last step the agent uses channel $\mathcal{E}_{SM \rightarrow SM}$ to extract work from the post-measurement state on $S$ and from the memory register $M$, thus ending up with two thermal states and making the protocol cyclic.
    \label{fig:scheme}}
\end{figure}

Let us now present a generic parameter estimation scheme that we will study throughout the rest of the paper. The protocol schematically
depicted in Fig.~\ref{fig:scheme} can be partitioned into three distinct steps: \emph{preparation}, which creates resource states that will be used during the estimation protocol, \emph{measurement} which effectively transfers information about the estimated parameter to the agent (via memory register)  and \emph{extraction} which brings all resource states back to equilibrium while extracting work. Below we present a detailed description of each of these steps.

\subsection{Estimation protocol stages}
\paragraph{Preparation.}
At the beginning of the protocol the agent creates two resource states: $\rho_S(x)$ and $\rho_M$. Information about a parameter $x \in \mathbb{R}$ will be encoded in system $S$, while system $M$ will serve as a classical memory register. For this purpose the agent draws for free maximally mixed states $\tau_S =  \mathbb{1}_{S} / |S|$ and $\tau_M = \mathbb{1}_M / |M|$ and using channel $\mathcal{P}_{SM \rightarrow SM}$ creates a joint state $\rho_{SM}$ in which system $S$ contains the desired dependence on the estimated parameter $x$ and $M$ is prepared in a ``reset'' memory state:
\begin{equation}
\label{eq:preparation}
    \mathcal{P}_{SM \rightarrow SM} \left[ \tau_S \ot \tau_M \right] = \rho_S (x) \ot \rho_M,
\end{equation}
where $\rho_M = \sum_j q_j \ket{j}\bra{j}_M$. We allow the initial memory state to be mixed, which is not a typical assumption in estimation problems as this reduces the information
content of the results stored in the register. However, since a mixed memory state requires less work to be prepared, it is not obvious that the agent should favour pure memory register over a noisy one when the available work is fixed. That is why we allow the register to be in a general state $\rho_M$.

As long as the parameter is unknown, part of the preparation step is in fact a parameter encoding process over which the agent does not have control. Still,  this does not change the analysis of the work cost. For example, if we consider a unitary parameter encoding process such as $\rho_S(x) =U_x \rho_S U_x^\dagger$, then we can easily split the probe state preparation and encoding stages and count only the work cost of preparing the state $\rho_S$ since the unitary acting on the system does not cost anything in the degenerate Hamiltonian framework. If the encoding is not unitary then we cannot perform such a split, since in principle the work cost may be different for different values of $x$ parameter. In this type of situations one should consider the whole preparation + encoding process as a single operation.

\paragraph{Measurement.}
In the next step the agent performs a general quantum measurement, which creates correlations between the measured system $S$ and a classical memory register $M$. The input state for this process is the state prepared during the preparation step, that is:
\begin{equation}
    \rho_{SM} = \rho_S(x) \ot \rho_M.
\end{equation}
A general quantum measurement corresponds to applying first a unitary $U$ which correlates $S$ with memory $M$ and then projecting the resultant state on one of the register eigenstates $\proj{k}_M$ for $k = 0, 1, \ldots |M| - 1$. The final state $\rho_{SM}^\prime$ can be written as:
\begin{eqnarray}
    \label{eq:8}
    \rho^\prime_{SM}
    &= \sum_{k = 0}^{|M|-1} \left[\mathbb{1}_S \ot \proj{k}_M \right] U \, \rho_{SM} \, U^{\dagger} \left[\mathbb{1}_S \ot \proj{k}_M \right] \\
    &= \mathcal{M}_{SM \rightarrow SM} \left[ \rho_{SM} \right],
\end{eqnarray}
where we labeled the measurement channel arising from unitary $U$ and projectors $\mathbb{1}_S \ot \proj{k}_M$ by $\mathcal{M}_{SM \rightarrow SM}$. Eq.~(\ref{eq:8}) can be written in simpler form as:
\begin{equation}
\label{eq:measurementmap}
    \mathcal{M}_{SM \rightarrow SM} \left[ \rho_{SM} \right] = \sum_{k,j = 0}^{|M|-1}  A_{k,j} \,  \rho_S(x) \, A_{k,j}^{\dagger} \ot \proj{k}_M,
\end{equation}
where operators $A_{k,j}$ are Kraus operators associated with map $\mathcal{M}_{SM \rightarrow SM}$ and given by: $A_{k,j} :=  \sqrt{q_j}\, \bra{k} U \ket{j}_M$, where $q_j$ are eigenvalues of the initial memory state $\rho_M = \sum_j q_j \ket{j}\bra{j}_M$.  The probability of obtaining outcome $k$ is then given by $p_k =  \tr \left[ \sum_j A_{k,j}^{\dagger} A_{k,j}\, \rho_S(x) \right] = \tr \left[M_k \, \rho_S(x) \right]$, where $M_k := \sum_j A_{k,j}^{\dagger} A_{k,j}$ is a POVM element associated with the measurement.
\paragraph{Work extraction.}
After the measurement agent ends up with a post-measurement state
 \begin{equation}
 \rho^\prime_{SM} = \sum_k  p_k \, \rho_{S,k}' \otimes \ket{k}\bra{k}_M,
\end{equation}
 where $\rho_{S,k}' = \sum_j A_{k,j} \rho_S A_{k,j}^{\dagger}/p_k$
 are the conditional post-measurement states. At this stage the  results of measurement are available to the agent and we assume that she utilizes it in an optimal way to estimate parameter $x$.

After that, we assume that the results are no longer relevant and the agent wants to extract as much work as possible while bringing the systems $S$ and $M$ back to equilibrium state. This assures that the whole estimation protocol becomes cyclic. Let $\mathcal{E}_{SM \rightarrow SM}$ denote the channel which extracts work from the joint state of system $SM$ after the measurement. We have that:
\begin{equation}
\label{eq:extractionmap}
    \mathcal{E}_{SM \rightarrow SM} \left[ \sum_k p_k \, \rho_{S,k}' \ot \proj{k}_M \right] = \tau_{S} \ot \tau_{M},
\end{equation}
where the output state $\tau_{S} \ot \tau_{M}$ is fixed, independently of the input.
\subsection{Precision of estimation}

\paragraph{Multishot.}
Determining measurement precision in the multi-shot regime is generally a well studied task and often can be done very efficiently due to the famous Cram{\'e}r-Rao bound \cite{Cramer1946}. Let us recall the notion of the Fisher information $F(x)$. For a probability distribution $p_k(x)$ depending on a parameter $x$, the \emph{Fisher information} is defined as:
\begin{equation}
	\label{eq:fisher}
    F(x) := \sum_{k} \frac{1}{p_k(x)}  \left( \frac{d p_k(x)}{ d x} \right)^2.
\end{equation}
In our case probabilities arise as a result of a POVM measurement $\{M_k\}$ acting on state $\rho_S(x)$, so that $p_k(x) = \tr \left[M_k\, \rho_S(x)\right]$. Intuitively speaking, Fisher information measures the average precision of estimation under observed data $k$. The Cram{\'e}r-Rao bound sets the lower limit of variance of any unbiased estimator $\tilde{x}(k)$  that assigns a value of the parameter based on the measured data, that is:
\begin{equation}
    \text{Var}(\tilde{x}) \geq \frac{1}{n F(x)} ,\quad \sqrt{n} \Delta \tilde{x} \geq \frac{1}{\sqrt{F(x)}},
\end{equation}
where $n$ is the number of repetitions of the experiment. Most importantly, the bound is saturable in the limit of large $n$, for example by application of the maximum likelihood estimator $\tilde{x}_{ML}$. More precisely, the maximum likelihood estimator when rescaled by $\sqrt{n}$ will asymptotically approach normal distribution centered at the rescaled true value of the parameter, with standard deviation given by $1/\sqrt{F(x)}$ so that $\sqrt{n}(\tilde{x}_{\t{ML}} - x) \approx \mathcal{N}(0, 1/F(x))$. Therefore, in the multi-shot scenario, Fisher information is the quantity that appropriately quantifies the performance of optimal estimation protocols.

Notice that we do not use here the concept of \emph{quantum} Fisher information \cite{Helstrom1976, Holevo1982}, where apart from optimization over the estimator, the measurement itself is also optimized to yield the best estimation performance. This is due to the fact that we will optimize estimation performance for a fixed work constraints. Therefore we may be forced to apply a sub-optimal measurement in order to reduce the amount of work invested in the protocol.

\paragraph{Single-shot.}
In statistics, and in particular in the estimation theory, the multi-shot regime is uncontroversial as different approaches to statistical inference lead to equivalent statements. Controversies, most notably between the frequentist and the Bayesian schools, become more pronounced while approaching the limit of fewer and fewer observations. The most extreme case is the single shot case, where we would like to provide a statement on the value of the parameter based on a single observation.

Arguably, the least controversial approach which avoids the problem of choosing a well justified prior which haunts the Bayesian approach,
and at the same time does not need to invoke the many-repetition argument necessary to justify the frequentist approach, is the concept of confidence intervals \cite{Schweder2016}. This approach recently found its application also in the quantum domain, see e.g. \cite{Blume2012}. In short, instead of providing a variance of the estimator (frequentist) or width of the  posteriori distribution (Bayesian), we first fix a certain confidence threshold $0 \leq \alpha \leq 1$ and  provide a corresponding region such that whatever the true value of the parameter was, the constructed region will contain it with probability at least $\alpha$. The natural way to construct such regions is based on the concept of the likelihood function.

Let $p_k(x)$ be the probability of observing an event $k$ (which may in principle represent results of many observations) given the true value of the parameter is $x$. The \emph{likelihood} function is nothing else than $p_k(x)$, but interpreted in a way that $x$ is now the varying argument of the function and $k$ is a fixed value. In other words, the likelihood function tells us what would be the probability of obtaining a given result $k$ (that we have actually observed, and hence is fixed) provided the true value of parameter was $x$. For a given measurement outcome $k$, one can define the \emph{log-likelihood ratio} function $\lambda_{k}(x)$ as:
\begin{equation}
    \lambda_k(x) := - 2 \log \left[ \frac{p_k(x)}{p_k[\tilde{x}_{\t{ML}}(k)]} \right],
\end{equation}
where $\tilde{x}_{\t{ML}}(k) = \t{argmax}_{x}\, p_k(x)$ is the maximum-likelihood estimator. The \emph{confidence region} for a given outcome $k$ and confidence level $\alpha$ is defined as:
\begin{equation}
    {\mathcal{R}}_{\alpha}(k) := \{ x \, | \, \lambda_k(x) < \lambda_{\alpha} \},
\end{equation}
where $\lambda_{\alpha}$ is a constant that depends only on the desired confidence level $\alpha$. In general, finding $\lambda_{\alpha}$  corresponds to solving the following equation for $\lambda_{\alpha}$:
\begin{equation}
	\label{eq:cregion}
    \max_{x} f(\lambda_{\alpha}|x) = 1 - \alpha,
\end{equation}
where $f(\lambda_{\alpha}|x) = \sum_{k: \lambda(x) > \lambda_{\alpha}} p_k(x)$, which can be easily found for simple models.

Here, we will adopt this approach when discussing single-shot quantum parameter estimation with  $p_k(x) = \tr \left[ M_k\, \rho_S(x) \right]$.  In particular, in the example in Section~\ref{sec:example}
we will choose the confidence level $\alpha = \t{Erf}(1/\sqrt{2}) \approx 0.68$, so that in the case of normal distribution the associated confidence interval will correspond to the $\pm \sigma$ interval. Consequently, if we applied the above philosophy to data obtained from a large number of experiment repetitions, then by the asymptotic normality arguments \cite{Schweder2016}, we would arrive at the confidence interval
$\mathcal{R}_k \approx  (\tilde{x}_{\t{ML}}(k) - 1/\sqrt{n F(x)}, \tilde{x}_{\t{ML}}(k) + 1/\sqrt{n F(x)})$. This interval coincides with the one-sigma error bars that we would obtain within the multi-shot scenario using the maximum-likelihood estimator,
$\tilde{x}_{\t{ML}}(k) \pm  1/\sqrt{n F(x)}$, and hence we can naturally relate the single-shot and the multi-shot figure of merits.


\section{Work cost of a quantum parameter estimation protocol}
\label{sec:workcostprotocol}
In this section we present both single and multi-shot work costs of the steps described in our estimation protocol.

\paragraph{Preparation.}
The preparation map $\mathcal{P}$, see Eq.~(\ref{eq:preparation}),  transforms a product state
$\tau_{S} \otimes \tau_M$ into a product state $\rho_S(x) \otimes \rho_M$.
According to (\ref{eq:workmulti}) the average work in the multi-shot regime reads:
\begin{eqnarray}
\langle w (\mathcal{P}) \rangle
	&= H(\tau_S \otimes \tau_M) - H(\rho_S\otimes \rho_M) \\ \nonumber
	&= H(\tau_S) - H(\rho_S) + H(\tau_M) - H(\rho_M),
\end{eqnarray}
where we made use of the entropy additivity property.

Since the channel is a \emph{fixed output} channel we may use a simplified formula (\ref{eq:workfixedout}) to find the work cost in the single-shot regime:
\begin{equation}
w(\mathcal{P}) = H_{\text{max}}(\tau_S) - H_{\text{min}}(\rho_S) + H_{\text{max}}(\tau_M) - H_{\text{min}}(\rho_M),
\end{equation}
where we used an analogous additivity property of the min and max-entropy \cite{Renner2008}.

\paragraph{Measurement.}
Multi-shot work cost of the measurement channel as defined in (\ref{eq:measurementmap}) reads:
\begin{eqnarray}
   \langle  w(\mathcal{M})\rangle  &= H(\rho_S \otimes \rho_M)  - H(\rho^\prime_{SM})  \\
   & = H(\rho_S) + H(\rho_M) - H(\{p_k\}) - \sum_k p_k H(\rho'_{S,k}).   \nonumber
\end{eqnarray}
The measurement channel is not the fixed-output channel and hence we cannot use the simplified formula for the single-shot work cost
(\ref{eq:workfixedout}). According to (\ref{eq:5}), the single-shot work cost reads:
\small
\begin{eqnarray}
   w(\mathcal{M}) = \log \Big\|\sum_{k = 0}^{|M|-1} \left[\mathbb{1}_S \ot \proj{k}_M \right] U \, \Pi_{\rho_S \otimes \rho_M}\, U^{\dagger} \left[\mathbb{1}_S \ot \proj{k}_M\right]\Big\|_\infty \leq 0,
\end{eqnarray}
\normalsize
where $\Pi_{\rho_S \otimes \rho_M}$ is a projection on the support of $\rho_S \otimes \rho_M$. If
$\rho_S$ and $\rho_M$ are full rank states then $\Pi_{\rho_S \otimes \rho_M} = \mathbb{1}_{SM}$, and the above expression yields
\begin{equation}
w^{\t{full rank}}(\mathcal{M}) = \log \norm{\mathbb{1}_{SM}}_\infty  = 0.
\end{equation}
In the other extreme case when both $\rho_S$ and $\rho_M$ are pure then
$\Pi_{\rho_S\otimes \rho_M} = \rho_S \otimes \rho_M$ and in this case
\begin{equation}
w^{\t{pure}}(\mathcal{M}) = \log \norm{\rho_{SM}^\prime}_\infty  = - H_{\t{min}}(\rho_{SM}^\prime).
\end{equation}
In the general case this work will take a value in between this two extreme cases so we can write
\begin{equation}
 w(\mathcal{M})  = - \eta H_{\t{min}}(\rho_{SM}^\prime), \quad 0 \leq \eta \leq 1,
\end{equation}
showing that this step will never cost us any work, but instead we may gain work by making the registers more noisy.

\paragraph{Work extraction.}
The work extraction channel $\mathcal{E_{SM \rightarrow SM}}$, see (\ref{eq:extractionmap}), is again a fixed-output channel that always yields
$\tau_S \otimes \tau_M$. Therefore, the average multi and single-shot work costs read respectively:
\begin{eqnarray}
    \langle w(\mathcal{E}) \rangle &= \sum_k p_k H(\rho_{S,k'}) + H(\{p_k\})   - H(\tau_S) - H(\tau_M), \\
    w(\mathcal{E}) & = H_{\t{max}}(\rho_{SM}^\prime)    - H_{\t{min}}(\tau_S) - H_{\t{min}}(\tau_M).
\end{eqnarray}
If $\rho_S$ and $\rho_M$ are full rank then $\rho_{SM}^\prime$ is full rank as well. In this case the single-shot work vanishes since
\begin{equation}
w^{\t{full rank}}(\mathcal{E}) = \log(|S|  |M|) - \log |D| - \log |M| = 0.
\end{equation}
Otherwise $w(\mathcal{E})$ may be negative and one may draw some deterministic work from this process with the maximal value $H_{\t{min}}(\tau_S) + H_{\t{min}}(\tau_M) = \log(|S|\cdot |M|)$ in an unlikely case when $\rho_{SM}^\prime$ is pure.

\paragraph{Total work cost and work credit}
Let us now analyze the combined work cost of all stages of the estimation protocol. In the multi-shot regime the total work vanishes, that is:
\begin{equation}
\langle w_{\t{total}} \rangle = \langle w(\mathcal{P}) \rangle  + \langle w(\mathcal{M}) \rangle  + \langle w(\mathcal{E}) \rangle  = 0,
\end{equation}
which corresponds to the fact that all steps are in principle reversible, and since the scheme is cyclic then no net work is consumed.

In the single-shot regime this is no longer the case as the total work cost reads:
\begin{eqnarray}
w_{\t{total}} &= w(\mathcal{P})  + w(\mathcal{M}) + w(\mathcal{E}) =  \\
&= H_{\t{max}}(\rho_{SM}^\prime) - \eta  H_{\t{min}}(\rho_{SM}^\prime) - H_{\t{min}}(\rho_S) - H_{\t{min}}(\rho_M),
\end{eqnarray}
where we used the fact that for maximally mixed states $H_{\t{min}}(\tau) = H_{\t{max}}(\tau)$. In case when $\rho_M$ and $\rho_S$ are full rank, the above formula simplifies to:
\begin{equation}
w^{\t{full rank}}_{\t{total}} = \log(|M| |S|) - H_{\t{min}}(\rho_S) - H_{\t{min}}(\rho_M),
\end{equation}
while for pure input registers we have:
\begin{equation}
\label{eq:pureinput}
w^{\t{pure}}_{\t{total}} =  H_{\t{max}}(\rho_{SM}^\prime) - H_{\t{min}}(\rho_{SM}^\prime) - H_{\t{min}}(\rho_S) - H_{\t{min}}(\rho_M).
\end{equation}
In the single-shot regime the total work cost will always be positive except for trivial cases e.g. $\rho_S = \tau_S$, $\rho_M = \tau_M$, or in the case when registers are pure but so is $\rho_{SM}^\prime$, which correspond to measurement scenarios yielding no information about the estimated parameter.

Even though the total work cost in the multi-shot regime vanishes, the estimation protocol requires that the agent first invests a cerain amount of work during the preparation stage and only in the course of further steps can she recover this work. We may therefore define a quantity called \emph{work credit} which represents the maximal amount of work that needs to be at agent's disposal at some stage of the protocol.
Since the measurement and work extraction stages always supply a net work to the agent, the work credit is in fact just the work invested during the preparation stage. Therefore in multi and single-shot regimes work credit takes the form:
\begin{eqnarray}
\langle w_{\t{credit}} \rangle & = H(\tau_S) - H(\rho_S) + H(\tau_M) - H(\rho_M),
 \\
w_{\t{credit}} & =  H_{\text{max}}(\tau_S) - H_{\text{min}}(\rho_S)+ H_{\text{max}}(\tau_M) - H_{\text{min}}(\rho_M).
\end{eqnarray}
Note that in the case of full rank input states, the single-shot work credit coincides with the total work cost. The above results are summarized in Table~\ref{tab:1} for clarity.
\begin{table}[t!]
   \bgroup
       \def\arraystretch{2}
  \resizebox{\columnwidth}{!}{%
    \begin{tabular}{|c|c|c|}
        \hline
         \textbf{work} & \textbf{single-shot} & \textbf{multi-shot}        \\
         \hline
         total       &  $H_{\t{max}}(\rho_{SM}^\prime) - \eta  H_{\t{min}}(\rho_{SM}^\prime) - H_{\t{min}}(\rho_S) - H_{\t{min}}(\rho_M)$ &  0 \\
         \hline
         credit & $H_{\text{max}}(\tau_S) + H_{\text{max}}(\tau_M) -  H_{\text{min}}(\rho_S) - H_{\text{min}}(\rho_M)$ & $H(\tau_S)  + H(\tau_M)  - H(\rho_S) - H(\rho_M)$ \\ \hline
    \end{tabular}
    }
    \egroup
    \caption{Single and multi-shot total work and work credit costs. Note that $H(\tau_S) = H_{\t{max}}(\tau_S) = \log |S|$, and similarly for the $\tau_M$ state. In case of full rank input states $\rho_S$ and $\rho_M$ the single-shot total work formula simplifies and becomes identical to the formula for work credit, since $\eta=0$ and $H_{\t{max}}(\rho_{SM}^\prime) = \log(|S| |M|)$. }
    \label{tab:1}
\end{table}

\section{Single qubit phase estimation example}
\label{sec:example}
In this section we illustrate the general consideration presented above using the paradigmatic phase estimation protocol in its most elementary single-qubit case. We will also consider the memory register to be a qubit to simplify the analysis. We do not expect this last assumption to impact the performance of estimation schemes as all phase estimation schemes require only projective measurements for optimal performance \cite{Demkowicz2015}, and since the system is a qubit a two-outcome measurement should suffice to find the optimal schemes. Clearly, in our case we deal with additional work constraints, and thus in general we may  be forced to consider a more general POVM measurement than a projective one. We will be able to model this by taking into account noisy initial memory register $M$, but for simplicity of presentation we will still
restrict ourselves to a two-outcome POVM measurement.

The phase is imprinted on the state prepared by the agent according to the standard unitary
\begin{equation}
\label{eq:phase_imp_unitary}
 \rho_S(\varphi) = V_\varphi \rho_S V_\varphi^\dagger, \quad V_{\varphi} = e^{i \sigma_z \varphi/2}.
\end{equation}
Since the the work cost of preparation depends only  on the mixedness of the state, without loss of generality we may assume that the state
$\rho_S(\varphi)$ is prepared on the equator of the Bloch sphere, as these states are most sensitive to phase changes:
 \begin{equation}
 \rho_S(\varphi)  = \frac{1}{2}\left(\mathbb{1}_2 + r \cos \varphi \cdot \sigma_x + r \sin \varphi \cdot \sigma_y \right),
 \end{equation}
where $r$ is the length of Bloch vector of the input state  and $\sigma_i$ for $i = x,y,z$ are the ordinary Pauli matrices. We assume that the initial state of $M$ can also be mixed, and without loss of generality (since the measurement is only determined by the choice of the entangling unitary $U$ appearing in the measurement map) we may assume that it points in the $z$-direction:
\begin{equation}
\rho_M = \frac{1}{2}\left(\mathbb{1}_2 +  m \sigma_z \right).
\end{equation}
Let us now describe the measurement itself.
    Without loss of generality, we choose the unitary $U$ from (\ref{eq:8}) in such a way that when the measurement is performed on a pure memory register (for example in state $\ket{1}\bra{1}_M$), then the corresponding POVM elements lie on the equator of the Bloch sphere at some longitudinal angle $\phi \in [0, 2 \pi]$---considering measurements outside the equatorial plane introduces only additional noise while providing no relevant information since the probe state is restricted to be in the equatorial plane only. However, since in general the memory register can be prepared in a mixture of states $\ket{0} \bra{0}_M$ and $\ket{1} \bra{1}_M$, then our POVM's are effectively mixtures of projective measurements. The unitary which correlates $S$ with $M$ can be written as:
\begin{equation}
\label{eq:unitary}
U = \sum_{j=0,1} \ket{\phi} \bra{\phi}_S  \ot \ket{j} \bra{0}_M +  \ket{\phi + \pi} \bra{\phi + \pi}_S  \ot \ket{j} \bra{1}_M,
\end{equation}
where  $\ket{\phi}_S = \frac{1}{\sqrt{2}}\left( \ket{0}_S + e^{i \phi}\ket{1}_S \right)$ and $\phi \in [0, 2\pi]$ is the azimuthal angle on the equator of the Bloch sphere. The associated POVM elements are given by:
\begin{equation}
    \label{eq:povm}
    M_0 = \frac{1}{2}\left[\mathbb{1}_2 +  m (\cos \phi \cdot \sigma_x + \sin \phi \cdot \sigma_y) \right], \qquad M_1 = \mathbb{1}_2 - M_0.
\end{equation}
The two possible measurement outcomes $k = 0$ and $k = 1$ result with probabilities $p_k = \tr \left[ \rho_S(\varphi) \, M_k \right]$, that is:
\begin{equation}
    p_0 = \frac{1}{2} \left(1 + m\, r \cos (\phi - \varphi) \right), \qquad p_1 = 1 - p_0.
\end{equation}
Let us now employ the results from two previous sections to determine the total work cost and the work credit of the whole estimation protocol both in the single-shot and multi-shot regimes.

\paragraph{Multi-shot.}
First notice that in the multi-shot case the work credit (per one qubit) becomes (see Table \ref{tab:1}):
\begin{eqnarray}
\label{eq:mscreditDEG}
\langle w_{\text{credit}} \rangle &= 2 - H\left(\rho_S(\varphi)\right) - H\left(\rho_M\right) \\ \nonumber
&= 2 - h\left(\frac{1+r}{2}\right) - h\left(\frac{1+m}{2}\right),
\end{eqnarray}
where $h(x)$ is the ordinary binary entropy, $h(x) := -x \log x - (1-x) \log (1-x)$. We see that the only relevant quantities here are the purity of the input state as determined by $r$ and the initial purity of the register as described by $m$. Let us now find the optimal estimation precision treating $\langle w_{\text{credit}} \rangle$ as a fixed quantity, thus constraining the allowed measurement outcomes. As we stated before, this can by done using the Cramer-Rao bound which relates Fisher information of quantum state with variance of measurement results performed on this state. Using (\ref{eq:fisher}) we can easily find Fisher information (per one qubit) to be equal to:
\begin{equation}
    F(\varphi) = \frac{m^2 \, r^2 \sin^2 (\phi - \varphi)}{1- m^2 r^2 \cos^2 (\phi - \varphi)},
\end{equation}
In the limit of large number of repetitions $n$ this allows to express the optimal estimation uncertainty as:
\begin{equation}
\Delta \tilde{\varphi}  =  \frac{1}{\sqrt{F(\varphi) n }}.
\end{equation}
We can now determine the optimal precision of measurement for a fixed work credit by solving the following optimization problem:
\small
\begin{equation}
    \label{eq:optMS}
    \max_{r, m, \phi} \, \{\,\, F(\varphi) \, \,| 0 \leq r \leq 1,\, 0 \leq m \leq 1, \,0 \leq \phi \leq 2\pi,\,w = \langle w_{\text{credit}} \rangle \}.
\end{equation}
Notice first that the work credit is independent on the measurement angle $\phi$ so that we can choose one which maximizes Fisher information for a given $r$, $m$ and $\varphi$. This happens when $\phi - \varphi = \pi/2$ which also renders Fisher information $F(\varphi)$ symmetric in $m$ and $r$. Thus it can be easily deduced that the optimal solution is obtained when $r = m$, for which we have the optimal Fisher information $F(\varphi) = r^4$  with the constraint that $r$ must be determined from the formula for the work credit $\langle w_{\text{credit}} \rangle = 2(1 - h((1+r)/2))$. Finally, the optimal precision under given work credit constraint in the multi-shot regime reads:
\begin{equation}
\sqrt{n} \Delta \tilde{\varphi} \geq \frac{1}{r^2}, \quad \t{where }  h[(1+r)/2)] = 1 - \langle w_{\text{credit}} \rangle/2.
\end{equation}

\paragraph{Single-shot.}
Let us now move to the single-shot regime. Using Table \ref{tab:1} we can easily compute the single-shot work credit as:
\begin{equation}
    w_{\text{credit}} =  \log \left(1+r\right) + \log \left(1+m\right).
\end{equation}
Note that in most situations the single-shot work credit is equal to the total single-shot work cost (see Table \ref{tab:1}). However, in the unlikely case when the prepared state of the system $S$ and memory register $M$ are both pure ($r = m = 1$) and the measurement angle happens to be equal to the estimated parameter $\phi = \varphi$ (so that $H_{\text{max}}(\rho_{SM}') = 0$ and the joint state $\rho_{SM}$ from (\ref{eq:pureinput}) remains unchanged, that is $\rho_{SM}' = \rho_{SM}$), the total single-shot work cost $w_{\text{total}}$ vanishes. We exclude this unphysical case from our further considerations.

Let us now find the size of the confidence intervals associated with this particular measurement scheme. The log-likelihood ratios for parameter $\varphi$ given outcomes $k= 0,\, 1$ are given by:
\begin{equation}
\lambda_k(\varphi) = -2 \log \left[ \frac{1+(-1)^k \cdot m \, r \cos (\phi - \varphi)}{1 + m \, r} \right].
\end{equation}
Using this and solving (\ref{eq:cregion}) allows us to write the confidence intervals for outcome $k$ as:
\begin{equation}    \hat{\mathcal{R}}_{\alpha}(k) = \{ \tilde{\varphi}_{\text{ML}}(k) - \delta \tilde{\varphi}, \quad \tilde{\varphi}_{\text{ML}}(k) + \delta \tilde{\varphi} \}
\end{equation}
where $\tilde{\varphi}_{\text{ML}}(k) = \t{argmax}_{\varphi} \,\, \lambda_k(\varphi) $ is the max-likelihood estimator and $\delta \tilde{\varphi}$ determines half of the size of our confidence region. Note that the regions for $k = 0$ and $k = 1$ have the same size so further we will not state the dependence on $k$ explicitly. It is important to emphasize that when determining confidence intervals one has to prepare for the worst case scenario and, since the true value of parameter $\varphi$ can be arbitrary and precision $\delta \tilde{\varphi}$ depends only on the relative difference between the two angles $\phi$, $\varphi$, we may assume without loss of generality that the measurement angle $\phi = 0$ and vary only $\varphi$.

\begin{figure}
    \centering
    \includegraphics[width = \textwidth]{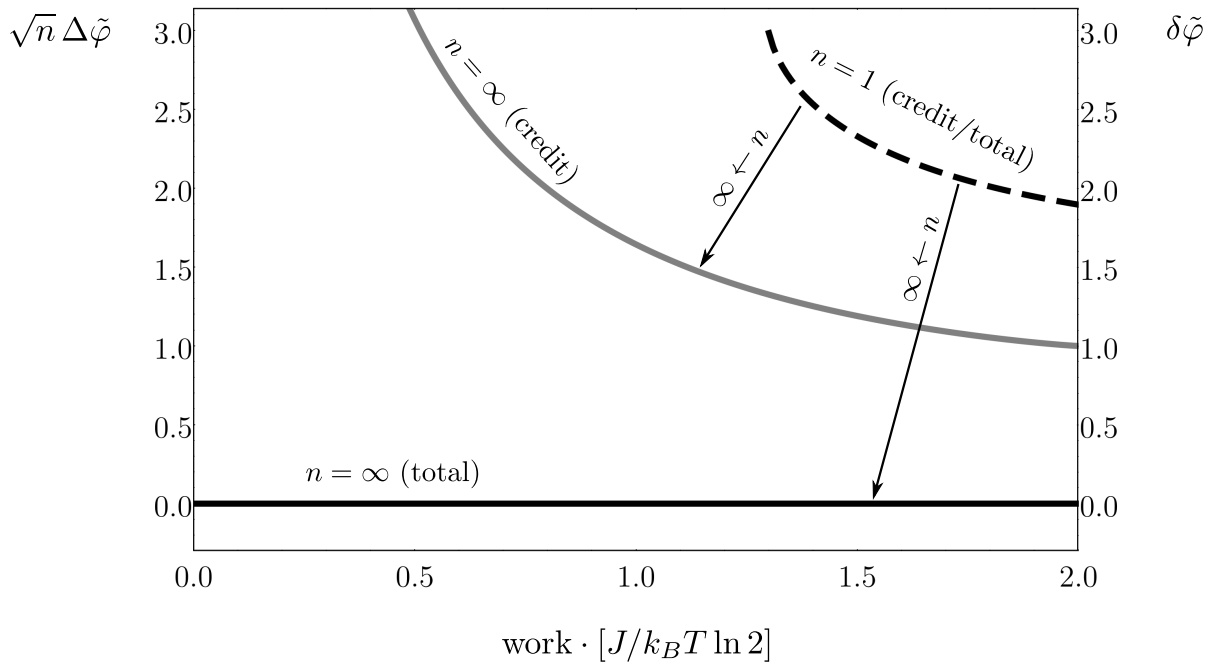}
    \caption{The maximal precision of estimation under a fixed work cost and fixed work credit. The dashed black line corresponds to the minimal size of the confidence interval $\delta \tilde{\varphi}$ for a fixed single-shot work credit as given by the solution of the optimization problem (\ref{eq:optSS}). In the case when agent prepares mixed input states (so that $r$ and $m$ are not equal to $1$) the single-shot work credit is equal to the single-shot total work cost). Solid grey line corresponds to the asymptotic case when $n = \infty$ and reprents the solution of the optimization problem (\ref{eq:optMS}). We see that in the many-copy limit size of the confidence interval $\delta \tilde{\varphi}$ effectively approaches the rescaled variance of the maximum-likelihood estimator $\sqrt{n} \Delta \tilde{\varphi}$ when the work credit is the constrained quantity. On the other hand, in the multi-shot regime the total work cost vanishes, thus making the protocol fully reversible, unlike in the single-shot case.}
    \label{fig:2}
\end{figure}

It can be easly shown that the size of the interval for confidence level $\alpha$ can be determined from:
\begin{equation}
    \delta \tilde{\varphi} = \arccos \left[ \frac{1-2\alpha}{m \cdot r} \right].
\end{equation}
Treating single-shot work credit $w_{\text{credit}}$ as a fixed quantity which effectively constrains the size of our confidence region as determined by $\delta \tilde{\varphi}$ allows us to state the following optimization problem:
\begin{equation}
\label{eq:optSS}
\min_{r, m} \quad \{\, \delta \tilde{\varphi} \, \,| \, 0 \leq r \leq 1, \,\, 0 \leq m \leq 1, \,\, w = w_{\text{credit}}\}  \\
\end{equation}
We can determine the optimal $\delta \tilde{\varphi}$ by noting that the goal function, as well as the constraint for the work credit, is symmetric in $r$ and $m$. The optimal parameters are thus $r = m = \sqrt{2^{w_{\text{credit}}}} - 1$, which yields the optimal confidence interval:
\begin{equation}
\delta \tilde{\varphi} =  \arccos \left[ \frac{1-2 \alpha}{(1- \sqrt{2^{w_{\text{credit}}}})^2} \right].
\end{equation}
In Fig.~\ref{fig:2} we plotted solutions of both (single and multi-shot) optimization problems.
As expected, the single-shot regime is much more demanding in terms of work consumed in order to perform an estimation protocol with the same
effective width of the confidence interval per probe system used.
\section{Extension to the non-degenerate Hamiltonian case}
\label{sec:nondeg}
Our analysis of the phase estimation task can be extended to scenarios which involve non-degenerate Hamiltonian of the probe system.
In order to stay consistent with the convention where the work was expressed in units $kT \ln 2$, we will express energies using dimensionless parameters $E_i$,  with the implicit assumption that they are again expressed in units $k T \ln 2$. Using this convention the dimensionless Hamiltonian of the system reads: $\mathcal{H}_X = \sum_{i} E_i \ket{i}\bra{i}_X$.
We will use recent results presented in \cite{Faist2018} concerning the minimal amount of work needed to execute channel $\mathcal{C}_{X \rightarrow X'}$, that unlike \cite{Faist2015}, takes into account the non-degenerate Hamiltonian case.
While the general formula for the minimal amount of work one has to supply (or can extract) to perform a given map $\mathcal{C}_{X \rightarrow X'}$ is quite involved, it simplifies in case where we require deterministic realization of the map and assume that the input states $\rho_X$ are full-rank. In this case the work cost takes the form:
\begin{equation}
    \label{eq:work_general}
    w(\mathcal{C})  = k T \ln 2 \cdot  \log \norm{ \Gamma_{\rho_{X'}}^{-1/2} \mathcal{C}_{X\rightarrow X'} \left[ \Gamma_{\rho_{X}} \right] \Gamma_{\rho_{X'}}^{-1/2} }_{\infty},
\end{equation}
where $\Gamma_{\rho_X} = \sum_{i \in\, \text{supp}\, \rho_X } 2^{-E_i} \ket{i}\bra{i}_X$ is a projection operator onto the support of the input state $\rho_X$  with Boltzmann weights $2^{-E_i}$ (or equivalently: $e^{- \beta E_i\, \cdot \, k T \ln 2} = 2^{-E_i}$) given to respective states $\ket{i}\bra{i}_X$. 
It is easy to see that when we set $\mathcal{H}_X= 0$ then the above formula reduces to $(\ref{eq:5})$.
 
 In the multi-shot limit (which requires considering an $\epsilon$-smooth version of (\ref{eq:work_general})), the work-cost of running channel $\mathcal{C}_{X \rightarrow X'}$ becomes:
\begin{equation}
\label{eq:workmulti2}
    \langle w(\mathcal{C}) \rangle = A(\rho_{X'}) - A(\rho_X),
\end{equation}
where $A(\rho_X) = \tr\left[ \mathcal{H}_X \rho_X \right] - kT\,  H(\rho_X)$, as could be expected, is the ordinary free energy. 

Similarly as in the case of trivial Hamiltonians, for channels with a fixed output the single-shot formula simplifies. Let $\mathcal{C}_{X \rightarrow X'}$ be a \emph{fixed-output} channel producing $\mathcal{C}_{X \rightarrow X'}[\rho_X] = \rho_{X'}$ irrespective of the input $\rho_X$, then we may write:
\begin{eqnarray}
\nonumber
    w(\mathcal{C})  &=  \log \norm{ \Gamma_{\rho_{X'}}^{-1/2} \mathcal{C}_{X\rightarrow X'} \left[ \Gamma_{\rho_{X}}/Z_{\rho_X} \right] \Gamma_{\rho_{X'}}^{-1/2} }_{\infty} + \log{Z_{\rho_X}} \\
    &=  \nonumber \log \norm{ \Gamma_{\rho_{X'}}^{-1/2} \rho_{X'} \Gamma_{\rho_{X'}}^{-1/2} }_{\infty} + \log{Z_{\rho_X}} \\
    &= \nonumber D_{\text{max}}(\rho_{X'} || \Gamma_{X'}) - D_{\text{min}}(\rho_{X} || \Gamma_{X}) \\
    \label{eq:work_fixed_out}
    &= A_{\text{max}}(\rho_{X'}) - A_{\text{min}}(\rho_{X}).
\end{eqnarray}
where $Z_{\rho_X} = \tr \, \left[\Gamma_{\rho_X}\right]$ is the partition function and $A_{\text{min}}(\rho_X) = D_{\min}(\rho_X||\Gamma_X) = - \log Z_{\rho_X}$  and $A_{\text{max}} = D_{\max}(\rho_X||\Gamma_X)= \log\norm{\Gamma_X^{-1/2} \rho_X \Gamma_X^{-1/2}}_{\infty}$ are the single-shot free energies first defined in \cite{Horodecki2013}, with $\Gamma_X =\sum_{i} 2^{-E_i} \ket{i}\bra{i}_X$, where the sum is not restricted to the support of $\rho_X$.

Let us now extend our study of the phase estimation protocol to the case when Hamiltonian of the probe system is no longer degenerate. We assume that the probe $S$ is now a qubit with Hamiltonian $H_S = \text{diag}(E_0, E_1)$ and without loss of generality, we choose $E_0 = 0$ and $E_1 = E$. Still, since memory register $M$ is just a logical device containing measurement results, we will again model it with a fully-degenerate Hamiltonian $\mathcal{H}_M = 0$. In the preparation step we start with a free (thermal) state on $SM$ given by $\tau_{SM} = \tau_S \ot \tau_M = \frac{1}{Z_S} 2^{-\mathcal{H}_S} \ot \frac{1}{2} \mathbb{1}_M $ with $Z_S = 1 + 2^{-E}$ being the partition function. The phase is imprinted on the state according to the same unitary as in the case from previous section (\ref{eq:phase_imp_unitary}), however now it is no longer true that preparing the state on the equator of the Bloch sphere costs the same amount of work as on any other parallel. That is why we need to consider a general qubit state at the preparation stage:
\begin{equation}
\rho_S(\varphi) = \frac{1}{2} \left(\mathbb{1}_2 + r \sin \theta \cos \varphi \cdot \sigma_x + r \sin \theta \sin \varphi \cdot \sigma_y + r \cos \theta \cdot \sigma_z \right),
\end{equation}
where again $r$ is the length of the Bloch vector, $\varphi$ is the phase to be determine and $\theta$ is the longitudinal angle determining distance from the equator. The initial state of register $M$ can again be mixed and using the same arguments as in the previous section we choose:
\begin{equation}
    \rho_M = \frac{1}{2} \left( \mathbb{1}_2 + m \cdot \sigma_z  \right).
\end{equation}
We shall now describe the measurement and its modifications with respect to the fully-degenerate case. We will use the same correlating unitary as in (\ref{eq:unitary}), thus the POVM elements associated with our measurement will be again given by (\ref{eq:povm}). The two possible outcomes $k = 0$ and $k = 1$ will result with probabilities:
\begin{equation}
    p_0 = \frac{1}{2} \left( 1 + m\, r \, \sin \theta \cdot \cos (\phi - \varphi) \right), \qquad p_1 = 1 - p_0.
\end{equation}
Let us now move to the calculation of the minimal amount of work necessary to carry out the estimation scheme.
\paragraph{Multi-shot.}
Similarly as in the degenerate case, multi-shot work costs can be computed simply by calculating differences in free energies (\ref{eq:workmulti2}) between subsequent states appearing in the protocol. Thus, the total multi-shot work cost of the whole scheme vanishes:
\begin{equation}
    \langle w_{\text{total}} \rangle = \langle w(\mathcal{P}) \rangle + \langle w(\mathcal{M}) \rangle + \langle w(\mathcal{E}) \rangle = 0.
\end{equation}
The multi-shot credit is given by:
\begin{eqnarray}
    \langle w_{\text{credit}} \rangle &= \langle w(\mathcal{P}) \rangle \\
    &= \left[ A(\rho_S(\varphi)) - A(\tau_S) \right] + \left[ H(\tau_M) - H(\rho_M)\right] \\
    &=\left[\frac{1}{2} E (1 - r \, \cos \theta) + \log Z_S - h\left(\frac{1 + r}{2} \right) \right] +  \left[1 - h\left(\frac{1 + m}{2} \right) \right],
\end{eqnarray}
and reduces to (\ref{eq:mscreditDEG}) when $E = 0$. The Fisher information can be straightforwardly calculated as:
\begin{equation}
F(\varphi) = \frac{m^2 r^2 \sin^2 (\phi- \varphi) \sin^2 \theta}{1 - m^2 r^2 \cos^2(\phi-\varphi) \sin^2 \theta}
\end{equation}
The optimal precision of measurement for a fixed work-credit can be evaluated by solving the following optimization problem:
\small
\begin{equation}
	\label{eq:optMSen}
      \max_{r, m, \phi, \theta} \, \{\,\, F(\varphi) \, \,| 0 \leq r \leq 1,\, 0 \leq m \leq 1, \,0 \leq \phi \leq 2\pi,\, 0 \leq \theta \leq \pi/2 ,\, w = \langle w_{\text{credit}} \rangle \}.
\end{equation}
By the same logic as before we may set $\varphi - \phi = \pi/2$, which again renders Fisher information symmetric in $r$ and $m$, however now the the work-credit depends on $r$ and $m$ in more complicated way. Moreover, the optimal solution depends now on the choice of energy gap $E$ and in general is always worse than in the case of a fully-degenerate Hamiltionian $\mathcal{H}_S = 0$.
\paragraph{Single-shot.}
Let us now address the total single-shot work cost and credit of estimating phase $\varphi$. First, note that the preparation map $\mathcal{P}$ which creates probe state $S$ and register $M$ along with encoding information about $\varphi$ in $S$, that is $\mathcal{P}[\tau_S \ot \tau_M] = \rho_S(\varphi) \ot \rho_M$ is a fixed-output channel and so we may apply formula (\ref{eq:workfixedout}) to get:
\begin{eqnarray}
    w(\mathcal{P}) &= \left[A_{\text{max}}(\rho_S(x)) - A_{\text{min}}(\tau_S)\right] + \left[ H_{\text{max}}(\tau_M) - H_{\text{min}}(\rho_M)\right] \\
    &= \log \norm{ \Gamma_{\rho_S}^{-1/2} \rho_S(x) \Gamma_{\rho_S}^{-1/2} }_\infty  + \log Z_S + \log(1+m) \\
    &= \log \left[ \lambda(r, \theta, E) \right] + \log(1+m),
\end{eqnarray}
where we labeled:
\begin{eqnarray*}
\lambda(r, \theta, E) = \frac{Z_S}{4} \, \Bigg[ 1 + z + 2^{E} (1-z) + \\
 \sqrt{(1+z)^2 + 2 \cdot 2^{E} (2r^2 -z^2 - 1) + 2^{2 E} (1-z)^2}\Bigg],
\end{eqnarray*}
where $z = r \, \cos \theta$. \\

\begin{figure}
    \centering
    \includegraphics[width = \textwidth]{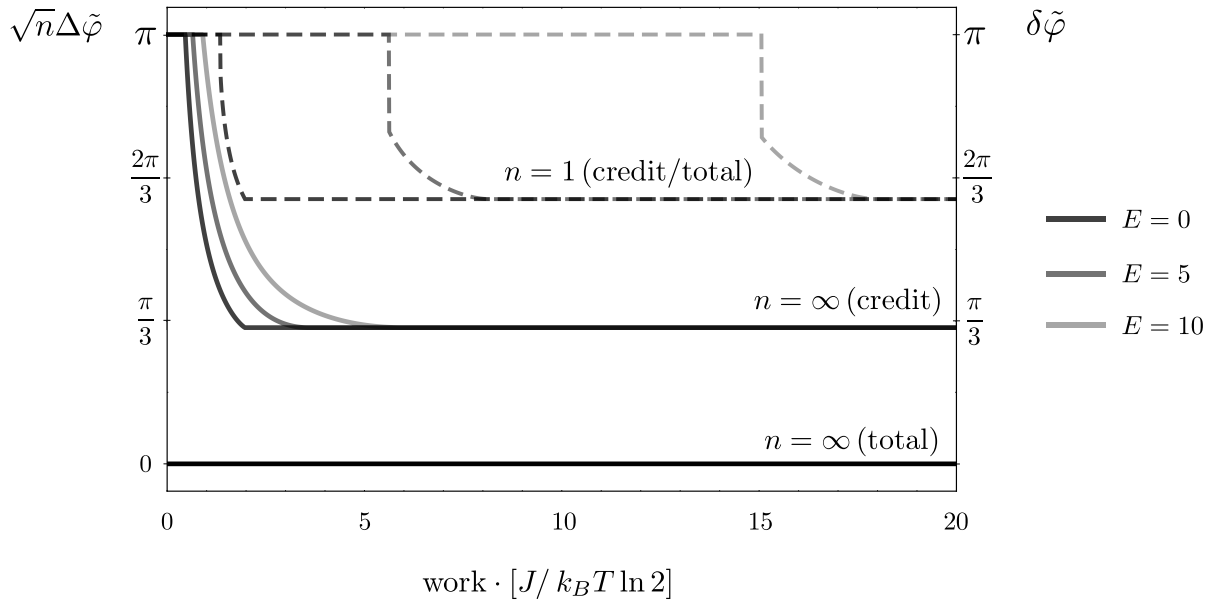}
    \caption{The maximal precision of estimation under a fixed work cost and credit when the probe system has a fixed energy gap $E$ (in units of $k T \ln 2$).
    The family of dashed lines correspond to the minimal sizes of confidence intervals $\delta \tilde{\varphi}$ for a fixed single-shot work credit and energy $E$ as given by the solution of the optimization problem (\ref{eq:optSSen}). Solid lines correspond to the asymptotic case when $n = \infty$ and represent curves obtained by numerically solving (\ref{eq:optMSen}). For simplicity of presentation we limited the $y-$axis to $\pi$, as this is the maximal possible variance (multi-shot) and the maximal possible confidence interval $\delta\tilde{\varphi}$ (single-shot) corresponding to the case when estimation does not provide any useful information about the phase. Setting $E = 0$ reproduces the same relation between work and precision as we described in Fig. (\ref{fig:2}). When the gap of the probe is positive, that is when $E \neq 0$, the agent must provide more work (as compared to the case when $E=0$) as now it requires additional work to prepare the state closer to the equator. Note that in the multi-shot regime the total work cost vanishes for all $E$ and thus the protocol becomes again fully reversible, just like in the degenerate case.}
    \label{fig:3}
\end{figure}

\noindent The work-cost of the measurement map (channel $\mathcal{M}$ defined in (\ref{eq:measurementmap})) can be calculated directly from (\ref{eq:work_general}). First, note that in our example the correlating unitary $U$  (see \ref{eq:unitary}) yields the following Kraus representation:
\begin{equation}
    A_{0j} = \sqrt{q_j} \, \ket{\phi} \bra{\phi}, \qquad A_{1j} =  \sqrt{q_j} \, \ket{\phi + \pi} \bra{\phi + \pi}.
\end{equation}
Let us now rewrite our Kraus operators as $A_{kj} = \sqrt{q_j} \ket{\phi_k} \bra{\phi_k}$, where $\phi_0 = \phi$ and $\phi_1 = \phi + \pi$ and note that for general (uncorrelated) input states $\rho_{SM} = \rho_S(\varphi) \ot \rho_M$ we have:
\begin{align}
\label{eq:work_M}
    \Gamma_{\rho_{SM}'}^{-1/2} \mathcal{M}(\Gamma_{\rho_{SM}}) \Gamma_{\rho_{SM}'}^{-1/2} &= \sum_{j \in \text{supp}  \, \rho_M} \sum_{k = 0}^{|M|-1} \left( \Gamma_{\rho_S'}^{-1/2} A_{kj} \, \Gamma_{\rho_S}^{1/2} \right) \left( \Gamma_{\rho_S}^{1/2} A_{kj}^{\dagger} \, \Gamma_{\rho_S'}^{-1/2} \right) \ot \ket{k}\bra{k}_M \\
    \nonumber
    &= \sum_{k = 0}^{|M|-1} \left( \sum_{j \in \text{supp}  \, \rho_M} q_j  \right) \,   \Gamma_{\rho_{S}'}^{-1/2} \ket{\phi_k}\bra{\phi_k} \Gamma_{\rho_S} \ket{\phi_k}\bra{\phi_k} \Gamma_{\rho_{S}'}^{-1/2} \ot \ket{k}\bra{k}_M \\
    \nonumber
    &= \sum_{k = 0}^{|M|-1} \Gamma_{\rho_{SM}'}^{-1/2} \ket{\phi_k} \bra{\phi_k} \Gamma_{\rho_{S}'}^{-1/2} \cdot \tr \left( \Gamma_{\rho_S} \ket{\phi_k}\bra{\phi_k} \right)\ot \ket{k}\bra{k}_M,
\end{align}
where we used the explicit form of $\ket{\phi_k} \bra{\phi_k}$ and by $\text{eig}(X)$ we mean the maximal eigenvalue of operator $X$. The work-cost $w(\mathcal{M})$ is given by the maximal eigenvalue of operator (\ref{eq:work_M}), which can also be written as:
\begin{align}
    w(\mathcal{M}) &= \max_k \,\,\,  \log \tr \left( \Gamma_{\rho_S} \ket{\phi_k} \bra{\phi_k}\right) \, + \, \log \text{eig} \left(    \Gamma_{\rho_{S}'}^{-1/2} \ket{\phi_k} \bra{\phi_k} \Gamma_{\rho_{S}'}^{-1/2}  \right) \\
    &= \log \tr \left( \frac{1}{2} \Gamma_{\rho_S} \right) + \log  \tr \left( \frac{1}{2} \Gamma_{\rho_S'}^{-1} \right) \\
    &= \log \Bigg( \frac{1}{2}  \sum_{i \in \text{supp} \, \rho_S} 2^{- E_i} \Bigg) + \log \left( \frac{1}{2}  \sum_{i \in \text{supp} \, \rho_S'} 2^{E_i} \right),
\end{align}
From the above it can be deduced that when input $\rho_S$ and output  $\rho_S'$ states are full-rank, then the associated work-cost $w(\mathcal{M})$ is given by:
\begin{equation}
    w^{\text{full rank}}(\mathcal{M}) = E + 2 (\log Z_S - 1).
\end{equation}
Note, however, that the work-cost of this step (excluding the unphysical case $r = m = 1$) is independent of $r$, $m$ and $\theta$ and depends solemnly on the energy gap $E$. Since our final aim is to optimize over the allowed parameters $(r, m, \theta)$ we can treat it as a constant and label: $c(E) =   E + 2 (\log Z_S - 1)$.
The last part of the protocol involves extracting work from post-measurement state $\rho_{SM}'$ using channel $\mathcal{E}$, and as such it is again a fixed-output channel. Thus we may again apply formula (\ref{eq:work_fixed_out}) to get:
\begin{align}
    w(\mathcal{E}) &= A_{\text{max}}(\tau_{SM}) - A_{\text{min}}(\rho_{SM}') \\
    &= \log \tr \left[ \Pi_{\rho_{SM}'} \, \Gamma_{\rho_{SM}'}\right] - \log Z_{SM} \leq 0.
\end{align}
Clearly, when $\rho_{SM}'$ is full-rank then we get $w^{\text{full rank}}(\mathcal{E}) = 0$, otherwise, if the post-measurement state $\rho_{SM}'$ is pure we get at most $w^{\text{pure}}(\mathcal{E}) = - \log Z_{SM}$, which by our convention is a work-yield. To summarize, we have the following total costs for the whole protocol:
\begin{eqnarray}
    w^{\text{full rank}}_{\text{total}} = w^{\text{full rank}}_{\text{credit}} = \log \left[ \lambda(r, \theta, E) \right] + \log(1+m) +c(E).
\end{eqnarray}
Let us now analyze the size of confidence intervals that we should attribute to our estimation scheme. The log-likelihood ratios for outcomes $k = 0$ and $k = 1$ take the following form:
\begin{equation}
    \lambda_k (\varphi) = -2 \log \left[ \frac{1 + (-1)^{-1} \cdot m r \sin \theta \cos (\phi - \varphi)}{1 + m r \sin \theta} \right]
\end{equation}
Just as in the $\mathcal{H}_S = 0$ case here we also find that both condifence intervals have equal sizes and thus we may focus on a specific outcome, say $k = 0$. Following the same arguments as before we find that the size of our confidence interval is given by:
\begin{equation}
    \delta \tilde{\varphi} = \arccos \left[ \frac{1 - 2 \alpha}{m\cdot r \cdot \sin \theta} \right].
\end{equation}
 We can now again treat $w_{\text{credit}}$ as a fixed quantity (and also $w_{\text{total}}$ as they are the same for full-rank inputs and outputs) and find a minimal size of confidence interval for this given work-cost. This can be written as the following optimization problem (for some fixed energy $E$):
 \begin{equation}
     \label{eq:optSSen}
\min_{r, m, \theta} \quad \{\, \delta \tilde{\varphi} \, \,| \, 0 \leq r \leq 1, \,\, 0 \leq m \leq 1, \,\, 0 \leq \theta \leq \pi/2, \,\, w = w_{\text{credit}}\}.
 \end{equation}
 This is an optimization problem where non-linearity is both in the goal function and the constraints. Note that here we cannot use symmetry arguments as we did in the fully-degenerate case as we should now also vary the longitudinal angle $\theta$. In Fig.~\ref{fig:3} we plotted numerical solutions of both (single and multi-shot) optimization problems. 
 
This solution is consistent with the intuition that when the probe system has a positive energy gap, then the agent must invest additional (as compared to the fully-degenerate case) amount of work to achieve the same precision of estimation. This becomes clear once we realize that the maximal precision of estimation can be achieved when the probe is prepared on the equator of the Bloch sphere. In the fully-degenerate case preparing the probe on the equator costs exactly as much work as on any other parallel, so without any additional expanses of work the agent can prepare the probe in that state (thus setting $\theta = \pi/2$) and then spend all of her work resources to suitably choose purity of the probe and the memory register by setting $(r, m)$. However, once the energy gap is positive, then this symmetry is broken and additional work must be provided in order to rotate the probe closer to the equator. This involves an additional cost as opposed to the previous case and forces the agent to prepare the probe at some suboptimal angle $\theta \neq \pi/2$ and suitably adjusted parameters $(r, m)$. 

Consider now the limiting case when energy gap of the probe is reasonably greater than $kT$, say $E = 10$, and the agent wants to obtain the best possible precision of estimation. It can be easly verified that, in the multi-shot regime, it is the cost of rotating the probe onto the equator ($ \approx 5\, k T$) which significantly dominates all other work-costs. On the other hand, when limited to the single-shot regime, it turns out that there are two main contributions to the total work-cost: the cost of rotating the probe onto the equator ($\approx 9\, kT$) and the cost of the measurement step itself ($\approx 8\, kT$). For further insight see Fig.~\ref{fig:3}. 

\section{Conclusions and Discussion}
\label{sec:conclusions}
The approach we advocate in this paper allows to study relations
between estimation precision in quantum metrological protocols and
their thermodynamic costs. While the Shannon entropy is the key
element to understand multi-shot asymptotic regime of this problem,
min and max-entropies as well as their smoothed version should be
used in any quantitative finite-shot analysis.

Our initial study may be developed further along many different
lines. First of all, we have not studied the
behaviour of smooth entropies in our protocol. This would be
necessary if we wanted to study the transition from single to multi-copy regime in a quantitative way. We would then need to fix a finite smoothing
parameter $\epsilon$, consider a finite probe version of the
protocol and optimize estimation precision under the work cost
determined by smooth entropic versions of the formulas derived in
this paper calculated on a finite number of copies of the considered quantum systems---we would expect the change in the curves plotted in Fig.~\ref{fig:2} as indicated by $n \rightarrow \infty$ arrows.

The other direction would be to go beyond the single qubit estimation
protocols and study multi-qubit phase estimation protocols with
possibly entangled inputs \cite{Adesso2016,Jozsa2003,Zhang2013,Joo2011,Sanders2012}. Even more challenging task would be to take into account decoherence effects. These are known on one hand
to fundamentally change the precision scaling of metrological
protocols
\cite{Demkowicz2012, Escher2011, Knysh2014, Jarzyna2017, Smirne2016, Nichols2016}, but on the other hand
they guarantee that single-shot considerations performed on multiple copies
are asymptotically equivalent to multi-shot results
\cite{Jarzyna2015, Demkowicz2015}.

\section{Acknowledgements}
We thank Michal Horodecki for fruitful discussions and Lucas Celeri and Paolo Giorda for their useful comments on the first draft of this paper. We also thank anonymous referees for their constructive comments. This work was supported by the (Polish) National Science Center grants No. 2016/22/E/ST2/00559 and 2017/27/N/ST2/01227.
\bibliographystyle{iopart-num}
\bibliography{citations}

\providecommand{\newblock}{}
\begin{thebibliography}{10}
\expandafter\ifx\csname url\endcsname\relax
  \def\url#1{{\tt #1}}\fi
\expandafter\ifx\csname urlprefix\endcsname\relax\def\urlprefix{URL }\fi
\providecommand{\eprint}[2][]{\url{#2}}

\bibitem{Landauer1961}
Landauer R 1961 {\em IBM J. Res. Dev.\/} {\bf 5} 183--191

\bibitem{Maxwell1871}
Maxwell J~C 2011 {\em {Theory of Heat}\/} (Cambridge University Press)

\bibitem{Bennett1982}
Bennett C~H 1982 {\em Int. J. Theor. Phys.\/} {\bf 21} 905----940

\bibitem{Helstrom1976}
Helstrom C~W 1976 {\em {Quantum Detection and Estimation Theory}\/} (Academic
  Press)

\bibitem{Holevo1982}
Holevo A~S 1982 {\em Probabilistic and Statistical Aspects of Quantum Theory\/}
  (North Holland, Amsterdam)

\bibitem{Giovannetti2006}
Giovannetti V, Lloyd S and Maccone L 2006 {\em Phys. Rev. Lett.\/} {\bf 96}(1)
  010401

\bibitem{Toth2014}
Toth G and Apellaniz I 2014 {\em J. Phys. A: Math. Theor.\/} {\bf 47} 424006

\bibitem{Demkowicz2015}
Demkowicz-Dobrzanski R, Jarzyna M and Ko\l{}ody\'{n}ski J 2015 {\em Quantum
  limits in optical interferometry\/} vol~60 (Elsevier)

\bibitem{Dorner2009}
Dorner U, Demkowicz-Dobrzanski R, Smith B, Lundeen J, Wasilewski W, Banaszek K
  and Walmsley I 2009 {\em Phys. Rev. Lett.\/} {\bf 102} 040403

\bibitem{Demkowicz2009}
Demkowicz-Dobrzanski R, Dorner U, Smith B, Lundeen J, Wasilewski W, Banaszek K
  and Walmsley I 2009 {\em Phys. Rev. A\/} {\bf 80} 013825

\bibitem{Mazzola2013}
Mazzola L, De~Chiara G and Paternostro M 2013 {\em Phys. Rev. Lett.\/} {\bf
  110}(23) 230602

\bibitem{Kolodynski2010}
Ko{\l}ody{\'n}ski J and Demkowicz-Dobrza{\'n}ski R 2010 {\em Phys. Rev. A\/}
  {\bf 82} 053804

\bibitem{Smirne2016}
Smirne A, Ko{\l}ody{\'n}ski J, Huelga S~F and Demkowicz-Dobrza{\'n}ski R 2016
  {\em Phys. Rev. Lett.\/} {\bf 116} 120801

\bibitem{Nichols2017}
Nichols R, Liuzzo-Scorpo P, Knott P~A and Adesso G 2018 {\em Phys. Rev. A\/}
  {\bf 98} 012114

\bibitem{Horodecki2013}
Horodecki M and Oppenheim J 2013 {\em Nat. Commun.\/} {\bf 4}

\bibitem{Goold2016}
{Goold} J, {Huber} M, {Riera} A, {del Rio} L and {Skrzypczyk} P 2016 {\em J.
  Phys. A: Math. Gen.\/} {\bf 49} 143001

\bibitem{Strasberg2017}
Strasberg P, Schaller G, Brandes T and Esposito M 2017 {\em Phys. Rev. X\/}
  {\bf 7} 021003

\bibitem{Faist2018}
Faist P and Renner R 2018 {\em Phys. Rev. X\/} {\bf 8} 021011

\bibitem{Skrzypczyk2014}
Skrzypczyk P, Short A~J and Popescu S 2014 {\em Nat. Commun.\/} {\bf 5}

\bibitem{Winter2016}
Winter A and Yang D 2016 {\em Phys. Rev. Lett.\/} {\bf 116} 120404

\bibitem{Oppenheim2002}
Oppenheim J, Horodecki M, Horodecki P and Horodecki R 2002 {\em Phys. Rev.
  Lett.\/} {\bf 89} 180402

\bibitem{Sparaciari2017}
Sparaciari C, Oppenheim J and Fritz T 2017 {\em Phys. Rev. A\/} {\bf 96}(5)
  052112

\bibitem{Linden2009}
Linden N, Popescu S, Short A~J and Winter A 2009 {\em Phys. Rev. E\/} {\bf 79}
  061103

\bibitem{Popescu2018}
Popescu S, Sainz A~B, Short A~J and Winter A 2018 {\em Philos Trans A Math Phys
  Eng Sci\/} {\bf 376}

\bibitem{Faist2015gpm}
Faist P, Oppenheim J and Renner R 2015 {\em New J. of Phys.\/} {\bf 17} 043003

\bibitem{Brandao2015}
Brand\~ao F~G~S~L and Gour G 2015 {\em Phys. Rev. Lett.\/} {\bf 115}(7) 070503

\bibitem{Guryanova2016}
Guryanova Y, Popescu S, Short A~J, Silva R and Skrzypczyk P 2016 {\em Nat.
  Commun.\/} {\bf 7} ncomms12049

\bibitem{Huber2015}
Huber M, Perarnau-Llobet M, Hovhannisyan K~V, Skrzypczyk P, Kl{\"o}ckl C,
  Brunner N and Ac{\'\i}n A 2015 {\em New J. of Phys.\/} {\bf 17} 065008

\bibitem{Chiribella2017}
Chiribella G and Yang Y 2017 {\em Phys. Rev. A\/} {\bf 96}(2) 022327
  \urlprefix\url{https://link.aps.org/doi/10.1103/PhysRevA.96.022327}

\bibitem{Lostaglio2015}
{Lostaglio} M, {Jennings} D and {Rudolph} T 2015 {\em Nature Communications\/}
  {\bf 6} 6383

\bibitem{Misra2016}
Misra A, Singh U, Bhattacharya S and Pati A~K 2016 {\em Phys. Rev. A\/} {\bf
  93} 052335

\bibitem{Micadei2013}
Micadei K, Serra R~M and C{\'e}leri L~C 2013 {\em Physical Review E\/} {\bf 88}
  062123

\bibitem{Scorpo2018}
Liuzzo-Scorpo P, Correa L~A, Pollock F~A, Gorecka A, Modi K and Adesso G 2018
  {\em New Journal of Physics\/} {\bf 20} 063009
  \urlprefix\url{http://stacks.iop.org/1367-2630/20/i=6/a=063009}

\bibitem{Erker2017}
Erker P, Mitchison M~T, Silva R, Woods M~P, Brunner N and Huber M 2017 {\em
  Phys. Rev. X\/} {\bf 7}(3) 031022
  \urlprefix\url{https://link.aps.org/doi/10.1103/PhysRevX.7.031022}

\bibitem{Korzekwa2016}
Korzekwa K, Lostaglio M, Oppenheim J and Jennings D 2016 {\em New J. Phys.\/}
  {\bf 18} 023045

\bibitem{Faist2015}
Faist P, Dupuis F, Oppenheim J and Renner R 2015 {\em Nat. Commun.\/} {\bf 6}

\bibitem{Feynman1998}
Feynman R 1998 {\em Feynman Lectures on Computation\/} (Addison-Wesley Longman
  Publishing Co., Inc.)

\bibitem{Szilard1964}
Szilard L 1964 {\em Behav. Sci.\/} {\bf 9} 301--310

\bibitem{Horodecki2005}
Horodecki M, Oppenheim J and Winter A 2005 {\em Nature\/} {\bf 436} 673--676

\bibitem{Renyi1961}
Renyi A 1961 On measures of entropy and information {\em Proceedings of the
  Fourth Berkeley Symposium...\/} (University of California Press) pp 547--561

\bibitem{Renner2008}
Renner R 2008 {\em International Journal of Quantum Information\/} {\bf 6}
  1--127

\bibitem{Renner2004}
Renner R and Wolf S 2004 Smooth renyi entropy and applications {\em
  International Symposium on Information Theory, 2004. ISIT 2004.
  Proceedings.\/} pp 233--242

\bibitem{Tomamichel2015}
Tomamichel M 2015 {\em Quantum Information Processing with Finite Resources:
  Mathematical Foundations\/} Springer Briefs in Mathematical Physics (Springer
  International Publishing) ISBN 9783319218915

\bibitem{Delrio2011}
Del~Rio L, {\AA}berg J, Renner R, Dahlsten O and Vedral V 2011 {\em Nature\/}
  {\bf 474} 61

\bibitem{Faist2016}
Faist P 2016 {\em {Quantum Coarse-Graining: An Information-Theoretic Approach
  to Thermodynamics}\/} (ETH)

\bibitem{Cramer1946}
Cramer H 1946 {\em Mathematical methods of statistics / by Harald Cramer\/}
  (Princeton University Press)

\bibitem{Schweder2016}
Schweder T and Hjort N~L 2016 {\em Confidence, Likelihood, Probability\/}
  (Cambridge Uni)

\bibitem{Blume2012}
Blume-Kohout R 2012 {\em arXiv preprint arXiv:1202.5270\/}

\bibitem{Adesso2016}
Adesso G, Bromley T~R and Cianciaruso M 2016 {\em J. Phys. A: Math. Theor.\/}
  {\bf 49} 473001

\bibitem{Jozsa2003}
Jozsa R, Koashi M, Linden N, Popescu S, Presnell S, Shepherd D and Winter A
  2003 {\em Quantum Information \& Computation\/} {\bf 3} 405--422

\bibitem{Zhang2013}
Zhang Y, Li X, Yang W and Jin G 2013 {\em Phys. Rev. A\/} {\bf 88} 043832

\bibitem{Joo2011}
Joo J, Munro W~J and Spiller T~P 2011 {\em Phys. Rev. Lett.\/} {\bf 107} 083601

\bibitem{Sanders2012}
Sanders B~C 2012 {\em J. Phys. A: Math. Theor.\/} {\bf 45} 244002

\bibitem{Demkowicz2012}
Demkowicz-Dobrza\'{n}ski R, Ko\l{}ody\'{n}ski J and {Gu{\c t}{\u a}} M 2012
  {\em Nat. Commun.\/} {\bf 3} 1063

\bibitem{Escher2011}
Escher B~M, de~Matos~Filho R~L and Davidovich L 2011 {\em Nature Phys.\/} {\bf
  7} 406--411

\bibitem{Knysh2014}
Knysh S~I, Chen E~H and Durkin G~A 2014 {\em arXiv preprint arXiv:1402.0495\/}

\bibitem{Jarzyna2017}
Jarzyna M and Zwierz M 2017 {\em Phys. Rev. A\/} {\bf 95} 012109

\bibitem{Nichols2016}
Nichols R, Bromley T~R, Correa L~A and Adesso G 2016 {\em Phys. Rev. A\/} {\bf
  94} 042101

\bibitem{Jarzyna2015}
Jarzyna M and Demkowicz-Dobrza\'{n}ski R 2015 {\em New J. Phys.\/} {\bf 17}
  013010

\end{thebibliography}
\end{document}